\documentclass[twocolumn,floats,floatfix,showpacs,amssymb,prd,superscriptaddress,nofootinbib]{revtex4-1}
\usepackage{graphicx, epsfig, amssymb} %include figure files
\usepackage{amsmath, amsfonts}
\usepackage{bm} %include bold math: \bm{} creates bold letters in math mode

\usepackage[linktocpage]{hyperref}
\usepackage[caption=false]{subfig}
\usepackage[usenames]{color}

\def\be{\begin{equation}}
\def\ee{\end{equation}}
\def\beq{\begin{eqnarray}}
\def\eeq{\end{eqnarray}}

\newcommand{\bea}{\begin{eqnarray}}
\newcommand{\eea}{\end{eqnarray}}
\newcommand{\ben}{\begin{enumerate}}
\newcommand{\een}{\end{enumerate}}
\newcommand{\bi}{\begin{itemize}}
\newcommand{\ei}{\end{itemize}}

\begin{document}

%\title{\large Awaking the vacuum in relativistic stars: the final state}
\title{\large The vacuum revealed: the final state of vacuum instabilities in
  compact stars}

\author{Paolo Pani}\email{paolo.pani@ist.utl.pt}
\affiliation{CENTRA, Departamento de F\'{\i}sica, 
Instituto Superior T\'ecnico, Universidade T\'ecnica de Lisboa - UTL,
Av.~Rovisco Pais 1, 1049 Lisboa, Portugal.}
\affiliation{Dipartimento di Fisica, Universit\`a di Cagliari, and INFN sezione di Cagliari, Cittadella
Universitaria 09042 Monserrato, Italy.}

\author{Vitor Cardoso} \email{vitor.cardoso@ist.utl.pt}
\affiliation{CENTRA, Departamento de F\'{\i}sica, 
Instituto Superior T\'ecnico, Universidade T\'ecnica de Lisboa - UTL,
Av.~Rovisco Pais 1, 1049 Lisboa, Portugal.}
\affiliation{Department of Physics and Astronomy, The University of Mississippi, University, MS 38677, USA.}

\author{Emanuele Berti} \email{berti@phy.olemiss.edu}
\affiliation{Department of Physics and Astronomy, The University of Mississippi, University, MS 38677, USA.}
\affiliation{California Institute of Technology, Pasadena, CA 91109, USA}

\author{Jocelyn Read} \email{jsread@relativity.phy.olemiss.edu}
\affiliation{Department of Physics and Astronomy, The University of Mississippi, University, MS 38677, USA.}

\author{Marcelo Salgado} \email{marcelo@nucleares.unam.mx}
\affiliation{Instituto de Ciencias Nucleares, Universidad Nacional Aut\'onoma
  de M\'exico, A.P. 70-543, M\'exico D.F. 04510, M\'exico}

%\date{\today} 

\begin{abstract}
Quantum fields in compact stars can be amplified due to a semiclassical
instability. This generic feature of scalar fields coupled to curvature may
affect the birth and the equilibrium structure of relativistic stars.  We
point out that the semiclassical instability has a classical counterpart,
which occurs exactly in the same region of the parameter space.  For negative
values of the coupling parameter the instability is equivalent to the
well-known ``spontaneous scalarization'' effect: the plausible end-state of
the instability is a static, asymptotically flat equilibrium configuration
with nonzero expectation value for the quantum fields, which is compatible
with experiments in the weak-field regime and energetically favored over
stellar solutions in general relativity.  For positive values of the coupling
parameter the new configurations are energetically disfavored, and the
end-point of the instability remains an open and interesting issue.
%The vacuum-driven instability can significantly modify the compactness,
%maximum mass and binding energy of compact stars. 
The vacuum instability may provide a natural mechanism to produce spontaneous
scalarization, leading to new experimental opportunities to probe the nature
of vacuum energy via astrophysical observations of compact stars.
\end{abstract}

\pacs{04.40.Dg, 04.62.+v, 95.30.Sf}

\maketitle

\date{today}
%%%%%%%%%%%%%%%%%%%%%%%%%%%%%%%%%%%%%%%%%%%%%%%%%%%%%%%%%%%%%%%%%%%%%%%%%%%%%%
%\section{Introduction}
%%%%%%%%%%%%%%%%%%%%%%%%%%%%%%%%%%%%%%%%%%%%%%%%%%%%%%%%%%%%%%%%%%%%%%%%%%%%%%
One of the great achievements of quantum field theory in curved spacetime is
Hawking's semiclassical prediction of black hole evaporation. This process is
made possible by the special nature of the vacuum in quantum field theory, but
it is extremely feeble by any astrophysical standards. 
%It is also characteristic of curved spacetimes with a very special property:
%the existence of an event horizon.
It was recently shown that the vacuum in strongly curved spacetimes might play
an important role even in the absence of horizons \cite{Lima:2010na,Barcelo},
so that the formation and evolution of relativistic stars could be affected by
semiclassical effects.

In particular, Lima, Matsas and Vanzella (henceforth LMV)
\cite{Lima:2010na} showed that the vacuum expectation value of nonminimally
coupled scalar fields can grow exponentially in relativistic stars.  
%LMV made some speculations on the final state of this instability, including
%collapse to a black hole geometry or an explosion (with consequent
%mass-shedding) that may lead to a stable stellar configuration with
%nonvanishing scalar field.
An understanding of the final state of the LMV instability is important: if
this final state supports nonvanishing scalar fields, the semiclassical
amplification of vacuum energy could have cosmological and astrophysical
implications.

In this paper we show that, for certain values of the coupling parameter, 
the most likely end-state of the instability is an asymptotically flat stellar solution with nonvanishing scalar field, which is compatible with gravitational experiments in the weak-field regime. We also point out that  these new solutions correspond to the well-known ``spontaneous scalarization'' phenomenon \cite{Damour:1993hw,Harada:1997mr}, and that they are {\it energetically favored} over stellar solutions in general relativity (GR). 
Our findings support the relevance of vacuum amplification scenarios. Vacuum amplification should be investigated carefully when devising strong-field tests of GR, for example in
the context of gravitational collapse and gravitational-wave emission.

%%%%%%%%%%%%%%%%%%%%%%%%%%%%%%%%%%%%%%%%%%%%%%%%%%%%%%%%%%%%%%%%%%%%%%%%%%%%%%
\section{Setup: static solutions}
%%%%%%%%%%%%%%%%%%%%%%%%%%%%%%%%%%%%%%%%%%%%%%%%%%%%%%%%%%%%%%%%%%%%%%%%%%%%%%
We look for static, spherically symmetric equilibrium solutions of the field
equations admitting a nonzero scalar field with metric
\be
ds^2=-f(r)dt^2+\left(1-2m(r)/r\right)^{-1}dr^2+r^2d\theta^2+r^2\sin^2\theta\,d\phi^2\,.\nonumber
\ee
%

%The nature of these solutions will provide strong evidence that they are a
%plausible end-state of the LMV instability. 
Following LMV~\cite{Lima:2010na}, we study a nonminimally coupled scalar field
in the presence of a perfect-fluid star in GR. We consider the action
\be S=\frac{1}{16\pi G}\, \int d^4x\sqrt{-g}
R + \int  d^4x\sqrt{-g} \, {\cal L}_m\,, 
\label{action}
\ee
where ${\cal L}_m={\cal L}_{\rm scalar}+{\cal L}_{\rm perfect \, fluid}$. The
Lagrangian for a scalar field with conformal coupling $\xi$ is given by
\be
{\cal L}_{\rm scalar}=
-\xi\,R\,\Phi^2-g^{\mu\nu}\Phi_{,\mu}\Phi_{,\nu}-\mu^2\Phi^2\,,
\ee
where $\mu$ denotes the scalar field mass. Here we set $\mu=0$; we will
consider the massive case in a follow-up paper.  For $\xi =1/6$, $\mu=0$ the
action is invariant under conformal transformations ($g_{\mu\nu}\rightarrow
\Omega^2g_{\mu\nu}\,,\Phi\rightarrow\Omega^{-1}\Phi$). For $\xi=0$, $\mu=0$
one recovers the usual minimally coupled massless scalar. The above Lagrangian
corresponds to a viable theory of gravity, passing all weak-field tests
\cite{Bronnikov:1973fh,Capozziello:2005bu}.

\begin{figure*}[htb]
\begin{center}
\begin{tabular}{cc}
\epsfig{file=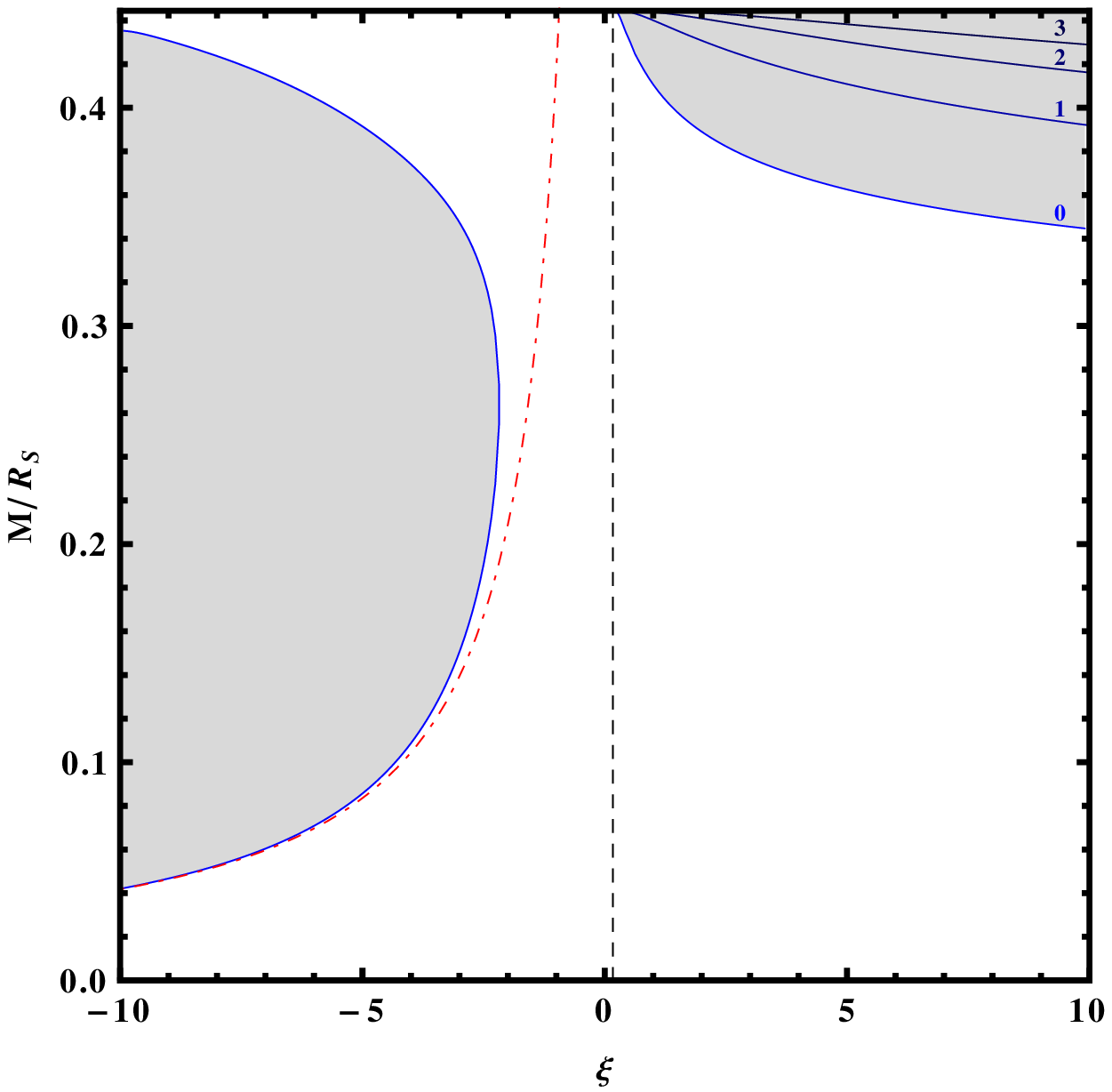,width=7.5cm,angle=0}&
\epsfig{file=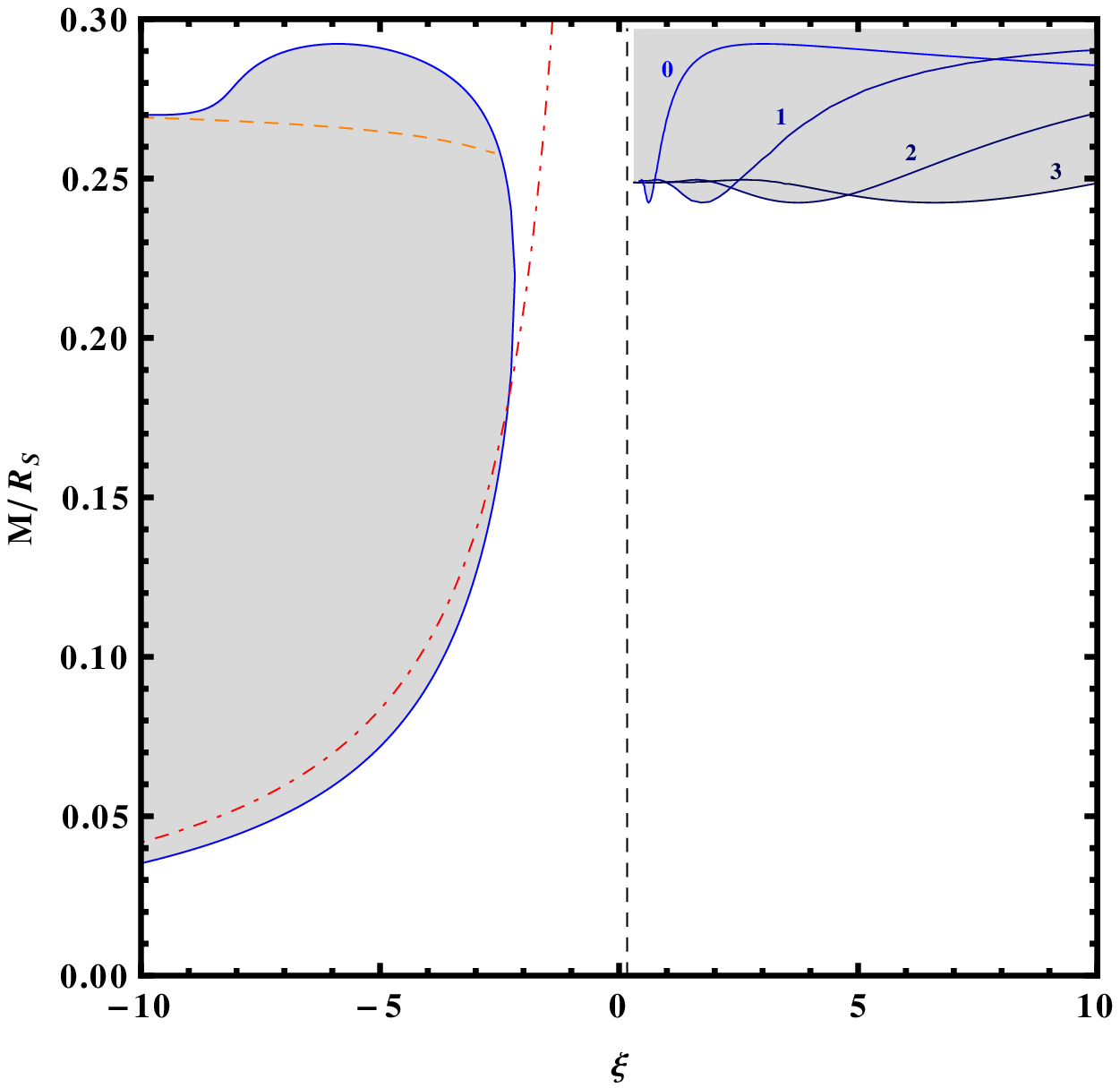,width=7.5cm,angle=0}
\end{tabular}
\caption{Left panel: Existence diagram for constant density stars, showing the
  values of the ratio $M/R_s$ of uniform density compact objects which support
  a dynamical scalar field. The dash-dotted line (red in the online version)
  corresponds to the analytical prediction in the Newton	ian limit,
  Eq.~(\ref{analytics}).  The vertical dashed line indicates the
  conformal-coupling value $\xi=1/6$. We present this plot in ``standard''
  geometrical units so that the compactness of our models can be easily
  compared to the maximum compactness for constant-density stars in GR:
  $M/R_s\leq 4/9\simeq 0.444$ (Buchdahl's limit). Modulo this trivial unit
  conversion, the shaded regions in the left panel match {\it exactly} the
  shaded regions in LMV's Fig.~1 (see main text).  In the upper-right region
  of the diagram, different critical lines correspond to ``excited''
  equilibrium configurations and the integers refer to the number $N$ of nodes
  in $\Phi(r)$ (see the main text).
  Right panel: Same for stars with a polytropic EOS. The dashed line (orange
  online) through the negative--$\xi$ region marks the configuration with
  maximum mass (i.e., the radial stability limit) for each
  $\xi$.\label{fig:diagram}}
\end{center}
\end{figure*}

The theory can be recast as a scalar-tensor theory of gravity in the Einstein
frame~\cite{Damour:1993hw} via the transformation
\be
g_{\mu\nu}\to(1-8\pi\xi\Phi^2)g_{\mu\nu}\,,\quad d\Phi\to\frac{\sqrt{1-8\pi\xi(1-6\xi)\Phi^2}}{1-8\pi\xi\Phi^2}d\Phi\,.\nonumber
\ee
In this form, both the classical counterpart of the LMV instability and its
final state have been thoroughly studied
\cite{Damour:1993hw,Harada:1997mr,Novak:1997hw}. 

We can explicitly construct static, asymptotically flat, spherically symmetric
solutions to the theory~(\ref{action}) with nonzero scalar field and verify
that our solutions match those found in
\cite{Damour:1993hw,Harada:1997mr,Novak:1997hw}, as follows. The equations of
motion following from the Lagrangian are (hereafter we set $c=G=1$)
\beq
G_{\mu\nu}&=&8\pi\,T_{\mu\nu}\,,\\
\nabla_{\alpha}\nabla^{\alpha} \Phi&=&(\mu^2+\xi\,R)\Phi\,,
\eeq
where
\begin{widetext}
\be
T_{\mu\nu}=\left(2\Phi_{,\mu}\Phi_{,\nu}-g_{\mu\nu}g^{\alpha\beta}\Phi_{,\alpha}\Phi_{,\beta}-\mu^2g_{\mu\nu}\Phi^2-2\xi\Phi^2_{,\mu;\nu}+2\xi\,g_{\mu\nu}\Phi^2_{,\alpha;\beta}g^{\alpha\beta}+T_{\mu\nu}^{\rm perfect\, fluid}\right)\left(1-16\pi\,\xi\Phi^2\right)^{-1}\,.
\ee
\end{widetext}
We consider perfect-fluid, spherically symmetric stars with energy density
$\rho(r)$ and pressure $P(r)$ such that $T^{\mu\nu}_{\rm perfect\,fluid}=
\left(\rho+P\right)u^\mu\,u^\nu+g^{\mu\nu}P$, where the fluid four-velocity
$u^\mu=(1/\sqrt{f},0,0,0)$. We specify some equation of state (EOS)
$P=P(\rho)$ and we impose regularity conditions at the center of the star,
i.e.
\be
m(0)=0\,,\quad \rho(0)=\rho_c\,,\quad \Phi(0)=\Phi_c\,,\quad \Phi'(0)=0 \,.
\ee
We also require continuity at the stellar radius $R_s$, defined by the
condition $P(R_s)=0$. We focus on two different stellar models: (i) the
constant density stars ($\rho={\rm const}$) studied by LMV, and (ii) the
polytropic model $\rho=n m_b+Kn_0 m_b (\Gamma-1)^{-1}(n/n_0)^\Gamma, P=Kn_0
m_b(n/n_0)^\Gamma$, with $\Gamma=2.34, K=0.0195, m_b=1.66\times 10^{-24}\,
{\rm g}$ and $n_0=0.1\,{\rm fm}^{-3}$ (this is the model that was considered
in Ref.~\cite{Damour:1993hw} in the context of spontaneous scalarization).  We
have checked that nuclear-physics based EOS models would yield qualitatively
similar results.
%A more detailed analysis will be presented elsewhere.

Relativistic stellar configurations in GR correspond to $\Phi_c=0$, so that
$\Phi=0$ everywhere. For each central density $\rho_c$, we used a shooting
method to search for nonzero values of $\Phi_c$ such that $\Phi(r)\to 0$ as
$r\to \infty$. 

%In this way we found a new class of asymptotically flat solutions with
%nonvanishing scalar field, which are described below.

%%%%%%%%%%%%%%%%%%%%%%%%%%%%%%%%%%%%%%%%%%%%%%%%%%%%%%%%%%%%%%%%%%%%%%%%%%%%%%
%%%Results
%%%%%%%%%%%%%%%%%%%%%%%%%%%%%%%%%%%%%%%%%%%%%%%%%%%%%%%%%%%%%%%%%%%%%%%%%%%%%%
For constant-density configurations, we find solutions with nonzero scalar
field in the shaded regions of the $(\xi\,,M/R_s)$ diagram shown in the left
panel of Fig.~\ref{fig:diagram}. The right panel of Fig.~\ref{fig:diagram}
shows that the qualitative features of the existence diagram are the same for
a polytropic EOS.
%From Fig.~\ref{fig:diagram}, larger deviations from pure GR are expected for
%high-compactness stars when $\xi>1/6$, and for intermediate compactness
%($M/R_s\sim0.2$) stars when $\xi\lesssim-2$.
These diagrams effectively reproduce previous results obtained many years ago
in the context of spontaneous scalarization (see
e.g.~\cite{Harada:1997mr}). It is remarkable that static solutions exist in
the {\it same} region where the LMV instability operates (cf.~Fig.~1 in
LMV). This provides strong evidence that these solutions (when they are
stable) represent a plausible final state of the instability.

In fact, the exact overlap between our own Fig.~\ref{fig:diagram} and Fig.~1
of LMV can be proved analytically.
Focus for simplicity on constant-density stars and massless scalar fields (but
our reasoning applies in general).  The critical lines in Fig.~1 of LMV
represent the curves where marginally stable modes exist. These modes are
zero-frequency solutions of Eq.~(4) in LMV, where the potential is given by
their Eq.~(6). On the other hand, the critical lines in our
Fig.~\ref{fig:diagram} represent the boundary of regions where spherically
symmetric, static solutions with nontrivial scalar field profiles cease to
exist. These are solutions of the Einstein-Klein-Gordon equations with
$\Phi=0$.
%In the limit of vanishing scalar field 
As $\Phi\to 0$ the spacetime becomes arbitrarily close to that of a
constant-density star and the Klein-Gordon equation reduces to Eq.~(4) in LMV,
with potential given by their Eq.~(6). Furthermore, the same boundary
conditions apply in both cases. Thus, the critical lines are obtained from the
{\em very same} equations and they are indeed coincident, not just similar.

\begin{figure*}[ht]
\begin{center}
\begin{tabular}{cc}
\epsfig{file=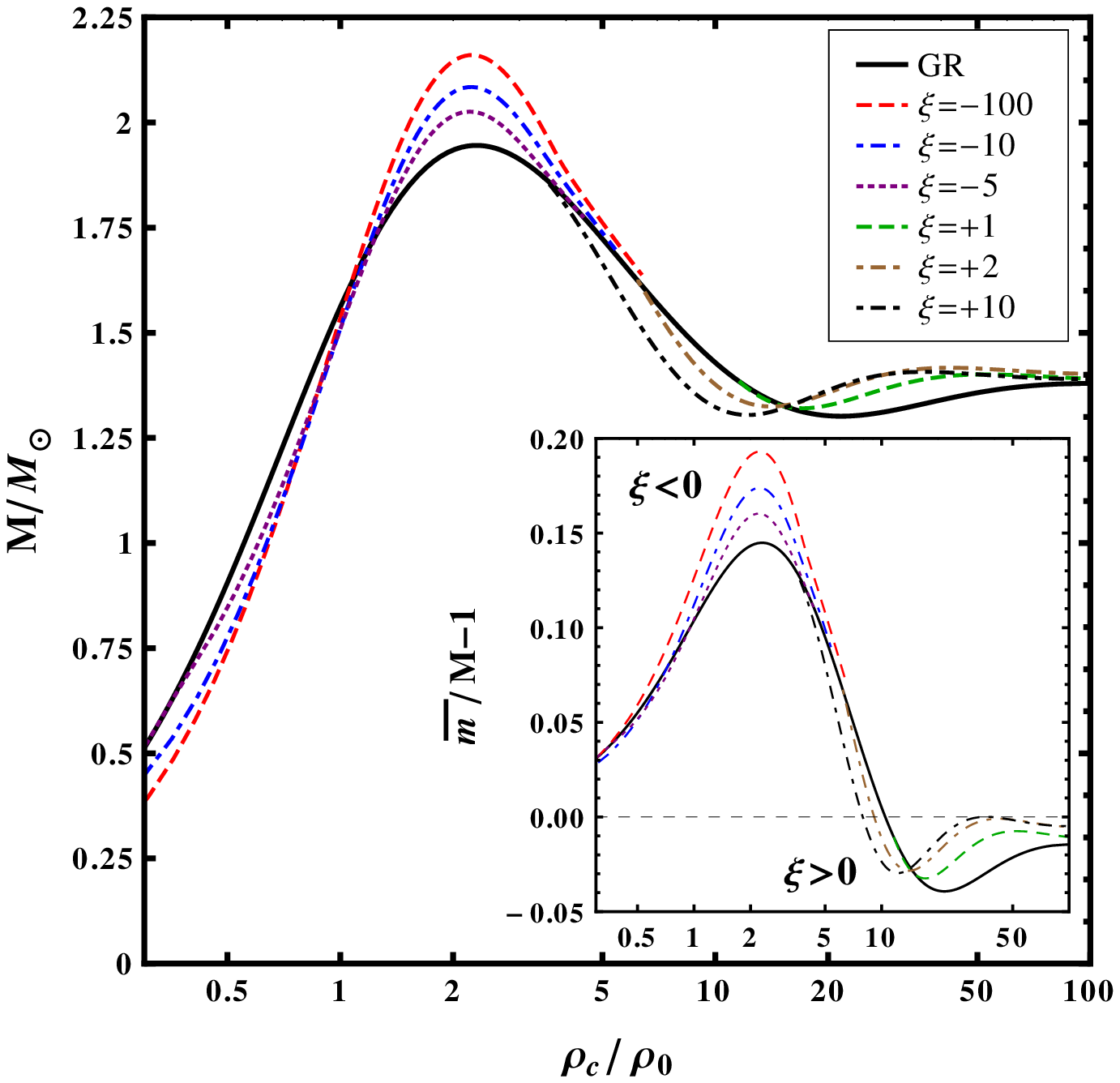,width=7.55cm,angle=0}&
\epsfig{file=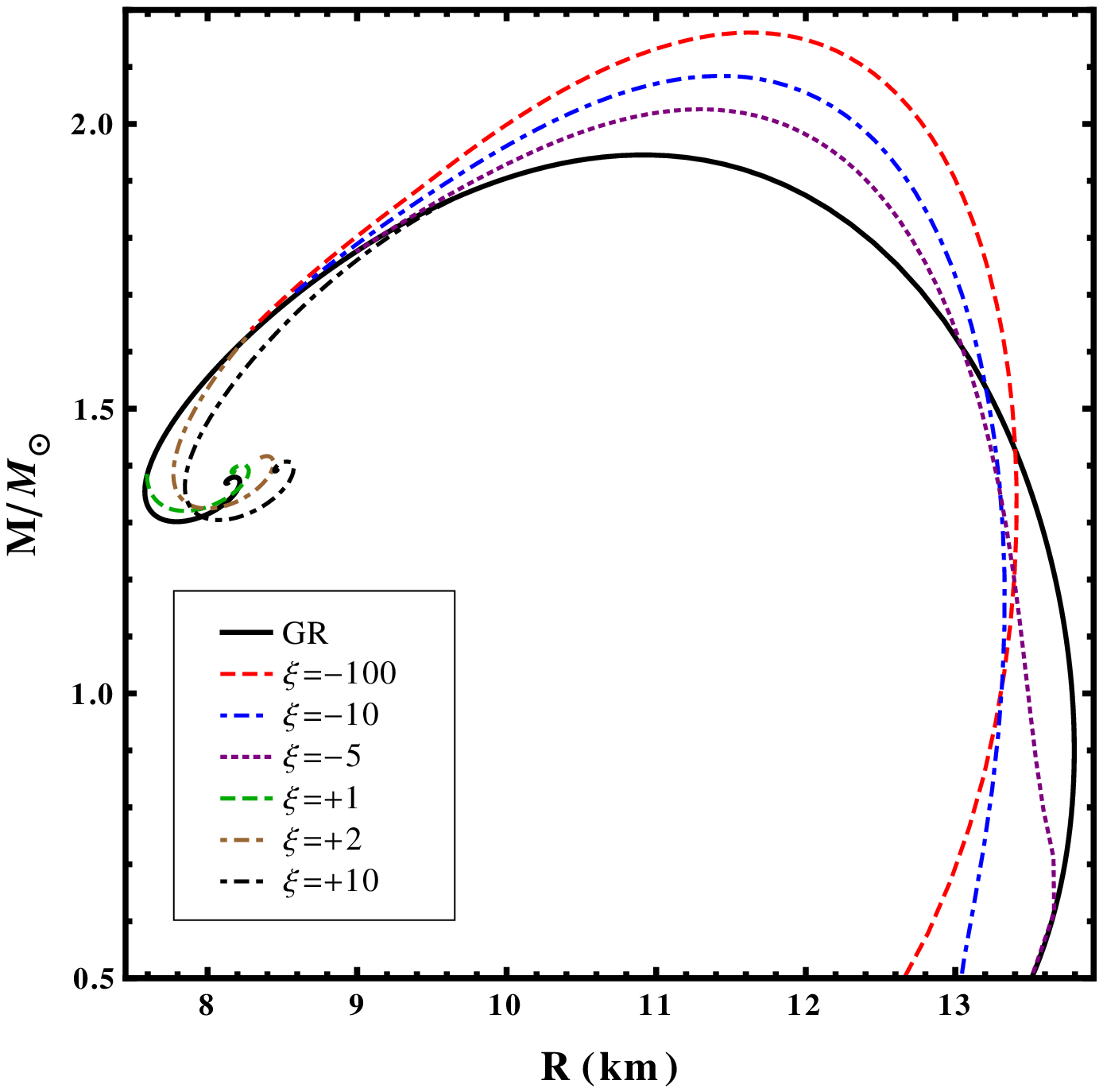,width=7.3cm,angle=0}
\end{tabular}
\caption{Left: Gravitational mass as a function of the central baryonic density
  $\rho_c/\rho_0$, where $\rho_0=8\times 10^{14}$~g/cm$^3$ is a typical
  central density for neutron stars. The inset shows the (normalized) binding
  energy as a function of $\rho_c/\rho_0$. Right: Gravitational mass as a function of the radius for different values of the coupling.
\label{fig:binding_energy}}
\end{center}
\end{figure*}

Quite interestingly, the LMV instability threshold can be found {\it
  analytically} in the Newtonian limit $M/R_s\ll 1$ (i.e., in the bottom left
corner of Fig.~\ref{fig:diagram}). The instability line defines the existence
of static solutions with a small but nonvanishing massless scalar field. The
relevant equation in this limit is $\Psi''-8\pi\,\xi( \rho-3P)\Psi=0$, where a
prime denotes a derivative with respect to $r$, and we use the ansatz for the
scalar field $\Phi=(\Psi/r)e^{-i\omega t}$ (i.e., we consider an s-wave).
Assuming that $\rho$ is constant in the stellar interior (this assumption
holds exactly for uniform-density stars and is a good approximation for most
EOSs), a regular solution at the origin and at infinity that is also
continuous at $R_s$ corresponds to
\be
24M \xi=-\pi^2\,R_s\,.\label{analytics}
\ee
As shown in Fig.~\ref{fig:diagram}, this prediction is in very good agreement
with the LMV results\footnote{The basic features of the instability in compact
  stars were understood by Ford (who studied unstable scalar fields as a
  possible mechanism to damp the effective value of the cosmological constant)
  as early as 1987 \cite{Ford:1987de}; see also \cite{Harada:1997mr}.}.

%%%%%%%%%%%%%%%%%%%%%%%%%%%%%%%%%%%%%%%%%%%%%%%%%%%%%%%%%%%%%%%%%%%%%%%%%%%%%%
\section{Mass and binding energy}
%%%%%%%%%%%%%%%%%%%%%%%%%%%%%%%%%%%%%%%%%%%%%%%%%%%%%%%%%%%%%%%%%%%%%%%%%%%%%%
The orbits of bodies far away from the star depend on the star's gravitational
mass $M$, shown as a function of the central baryonic density $\rho_c=m_b
n(0)$ in the left panel of Fig.~\ref{fig:binding_energy}. In GR, maxima of
this curve correspond to marginally stable equilibrium configurations, and all
solutions after the first maximum are unstable to radial perturbations in the
polytropic case (see e.g.~\cite{Shapiro:1983du}). %
The baryonic mass of the configuration,
\be \bar{m}=m_b\int d^3x\sqrt{-g}u^0 n(r)\,, \ee
corresponds to the energy that the system would have if all baryons were
dispersed to infinity.  The normalized binding energy $E_b/M=\bar{m}/M-1$ is
plotted in the inset of the left panel in Fig.~\ref{fig:binding_energy}.
For bound (not necessarily stable) configurations, $E_b>0$.
%$\bar{m}>M$.  

%Models
%with positive and negative coupling constant behave quite differently, and
%we find it convenient to discuss them separately.

%%%%%%%%%%%%%%%%%%%%%%%%%%%%%%%%%%%%%%%%%%%%%%%%%%%%%%%%%%%%%%%%%%%%%%%%%%%%%%
\subsection{Negative $\xi$}
%%%%%%%%%%%%%%%%%%%%%%%%%%%%%%%%%%%%%%%%%%%%%%%%%%%%%%%%%%%%%%%%%%%%%%%%%%%%%%
%%%%%%%%%%%%%%%%%%%%%%%%%%%%%%%%%%%%%%%%%%%%%%%%%%%%%
%\subsubsection{Negative $\xi$}
%%%%%%%%%%%%%%%%%%%%%%%%%%%%%%%%%%%%%%%%%%%%%%%%%%%%%
When $\xi\lesssim-2$, in the shaded region of Fig.~\ref{fig:diagram} we found
only a single solution coupled to a nontrivial scalar field. As shown in the
left panel of Fig.~\ref{fig:binding_energy}, the critical central density for
these solutions is roughly the same as in GR, but they have larger maximum
mass than their GR counterparts ($\approx 10\%$ larger for $\xi\lesssim-5$).
Similar arguments to GR indicate that models after the first maximum in
$M(\rho_c)$ (lying above the dashed line in the negative--$\xi$ region of the
right panel of Fig.~\ref{fig:diagram}) are unstable.

The inset of the left panel in Fig.~\ref{fig:binding_energy} shows that
similar deviations occur also for the binding energy. For a given $\bar{m}$,
the binding energy is \emph{higher} than in GR, so these configurations are
energetically favored over stellar solutions in GR.  As shown in the right
panel of Fig.~\ref{fig:binding_energy}, for a fixed EOS the scalar field can
sensibly modify the mass-radius relation. These modifications depend on $\xi$,
but they can be of order 10\% or more, and as such they could be
observable. Present neutron star observations give constraints on the
mass-radius relationship (e.g. \cite{Steiner:2010fz,Ozel:2010}), and
electromagnetic observations of binaries containing X-ray pulsars could in
principle constrain the binding energy as well
\cite{Alecian:2003ez,Spyrou:2001}.  A recent high-mass neutron star
observation (\cite{Demorest:2010}) also rules out many EOS models in GR.  If
astrophysical measurements are not compatible with any realistic EOS in pure
GR, we may be able to constrain the coupling constant $\xi$ and to probe the
occurrence of vacuum instabilities in astrophysical environments.

%%%%%%%%%%%%%%%%%%%%%%%%%%%%%%%%%%%%%%%%%%%%%%%%%%%%%%%%%%%%%%%%%%%%%%%%%%%%%%
\subsection{Positive $\xi$}
%%%%%%%%%%%%%%%%%%%%%%%%%%%%%%%%%%%%%%%%%%%%%%%%%%%%%%%%%%%%%%%%%%%%%%%%%%%%%%
%%%%%%%%%%%%%%%%%%%%%%%%%%%%%%%%%%%%%%%%%%%%%%%%%%%%%
%\subsubsection{Positive $\xi$}
%%%%%%%%%%%%%%%%%%%%%%%%%%%%%%%%%%%%%%%%%%%%%%%%%%%%%
For $\xi>1/6$ the situation is more complicated. Depending on the compactness
$M/R_s$ and on the coupling constant $\xi$, we have found a hierarchy of new
solutions, which can be labeled by the number $N$ of radial nodes of the
scalar field. In general, the larger $N$, the lower the binding energy, so
high-$N$ solutions can be thought of as ``excitations'' of the (energetically
favored) ground-state configuration.  Critical curves bounding the shaded
existence region of these solutions are plotted in the upper-right corner of
the two panels of Fig.~\ref{fig:diagram}.

%These solutions are energetically favored only at central densities larger
%than that corresponding to the maximum mass, so they should be unstable. 
The inset of Fig.~\ref{fig:binding_energy} shows that ground-state
configurations actually have a {\it negative} binding energy for
$\rho_c>\rho_\text{crit}\sim 6\times 10^{15}$ g/cm$^3$.
%For $\xi\to+\infty$
%the second local maximum approaches $E_b=0$, i.e., the configuration is
%only marginally bound.  
When $\rho_c<\rho_\text{crit}$, we only found solutions with binding energy
{\it lower} than in GR, so GR solutions are energetically favored in this
region of parameter space.
The fact that negative-$\xi$ solutions are more stable than positive-$\xi$
solutions is consistent with the interpretation of $\xi$ as an ``effective
gravitational constant'' proposed by van der Bij and Gleiser
\cite{vanderBij:1987gi} in their study of boson stars with nonminimal
coupling.

It is reasonable to conjecture that the top-right region of
Fig.~\ref{fig:diagram} does not correspond to stable ``stars'' with
nonvanishing scalar field.  The final state of the instability in this case is
an interesting topic for numerical simulations~\cite{Novak:1997hw} (see
also \cite{Balakrishna:2006ru} and references therein). Note that no-hair
theorems for positive $\xi$ (in particular $\xi\geq 1/2$) have been proven
years ago \cite{Mayo:1996mv,Saa:1996aw} and supported by numerical searches
\cite{Pena:2001sz}. This may imply that either stars with $\xi>1/6$ evolve to
a black hole solution in pure GR, or that they shed enough mass to leave the
forbidden region and become a star in pure GR again.

%%%%%%%%%%%%%%%%%%%%%%%%%%%%%%%%%%%%%%%%%%%%%%%%%%%%%%%%%%%%%%%%%%%%%%%%%%%%%%
\section{Extensions}
%%%%%%%%%%%%%%%%%%%%%%%%%%%%%%%%%%%%%%%%%%%%%%%%%%%%%%%%%%%%%%%%%%%%%%%%%%%%%%
The LMV mechanism, like other strong-field effects in scalar-tensor
theories \cite{Damour:1992we,Damour:1993hw}, relies on a specific coupling of
the scalar field with matter. We found that an instability is {\em not}
present for ``minimally-coupled'' Weyl fermions or Maxwell fields,
independently of the EOS\footnote{The electromagnetic case can be treated as
  described in Ref.~\cite{ruffini}. This leads to an equation of the form
  ${d^2\Psi}/{dr_*^2}+\left(\omega^2-V\right)\Psi=0$ for both polarization
  states, where we defined $dr/dr_*=\sqrt{f(1-2m/r)}$, and $V=f l(l+1)/r^2$.
  Massless neutrino fields yield a similar equation with potential
  $V=(2r^2)^{-1}[2k f(r)\left(k+\sqrt{1-2m/r}\right)-kr\sqrt{1-2m/r}f']$,
  where $k$ is a separation constant \cite{Brill:1957fx}. The potential for
  Maxwell perturbations is positive-definite, and therefore no instabilities
  can arise. For neutrinos we were not able to prove stability in general, but
  we did verify that the two specific models for the EOS discussed in this
  paper lead to positive-definite potentials.}. 
%Nonminimal couplings to
%curvature generically lead to ``spontaneous scalarization''. A generalization
%of this study to other theories is underway \cite{future}.

We have not explicitly shown that the equilibrium configurations reported here
arise as a result of nonlinear time evolutions of GR solutions with a ``seed''
scalar field; however, strong indications that this should be the case come
from numerical studies by Novak in the context of spontaneous scalarization
\cite{Novak:1997hw}. It would be interesting to perform similar simulations
for $\xi>1/6$.
%The larger potential binding energy stored in the star for nonminimally
%coupled theories may produce large amounts of gravitational radiation during
%collapse.
Other possible extensions of our work concern the investigation of slowly and
rapidly rotating stellar models and of their oscillation frequencies (see
e.g. \cite{Sotani:2005qx}).

%New stationary configurations coupled to scalar fields can potentially exist.
%Furthermore, the different structure of stellar configurations in these
%theories can produce large (and potentially observable) modifications of
%their The study of nonradial oscillations in this context is difficult but
%important: the direct detection of gravitational waves from compact objects
%could open an experimental laboratory to probe the field equations of
%gravitation.

%%%%%%%%%%%%%%%%%%%%%%%%%%%%%%%%%%%%%%%%%%%%%%%%%%%%%%%%%%%%%%%%%%%%%%%%%%%%%%
\section{Conclusions}
%%%%%%%%%%%%%%%%%%%%%%%%%%%%%%%%%%%%%%%%%%%%%%%%%%%%%%%%%%%%%%%%%%%%%%%%%%%%%%
We have reconsidered a generic class of theories where a scalar field is
nonminimally coupled to the Ricci scalar, that were recently shown to give
rise to a semiclassical instability.  
We have pointed out an interesting relation between the semiclassical instability and the spontaneous scalarization effect in classical scalar-tensor theories.
For certain values of the
coupling parameter the scalar field can leave observable imprints on the
equilibrium properties of relativistic stars. Our main finding is that the LMV
instability may provide a ``natural'' seed mechanism to produce spontaneous
scalarization, reinforcing the relevance of previous studies of compact stars
in scalar-tensor or $f(R)$ theories of gravitation (see
e.g.~\cite{Damour:1992we,Damour:1993hw,Babichev:2009fi}).

We stress that corrections to GR due to scalar fields are a {\it generic}
feature of a large class of unification theories. Our work suggests that
strong-field modifications to GR compatible with weak-field tests may be
astrophysically viable, with potentially observable consequences in the
structure of compact stars. It will be important to explore the implications
of vacuum amplification mechanisms for tests of strong-field gravity in
compact objects (see e.g. \cite{Psaltis:2008bb} and references therein).

%More exotic possibilities exist. For example ``boson stars'' are compact
%configurations of self-gravitating, \emph{charged} and/or \emph{massive}
%scalar fields. In the absence of ordinary matter, they require
%self-interactions~\cite{Colpi:1986ye} or large non-minimal
%couplings~\cite{vanderBij:1987gi} in order to have a mass comparable to the
%Chandrasekhar limit. We are working to understand how boson stars are affected
%by the LMV mechanism, and their connection with the solutions we found in
%this work.

%%%%%%%%%%%%%%%%%%%%%%%%%%%%%%%%%%%%%%%%%%%%%%%%%%%%%%%%%%%%%%%%%%%%%%%%%%%%%%
%\section{Acknowledgements}
{\bf \em Acknowledgements.}
%%%%%%%%%%%%%%%%%%%%%%%%%%%%%%%%%%%%%%%%%%%%%%%%%%%%%%%%%%%%%%%%%%%%%%%%%%%%%% 
We thank Yanbei Chen for useful discussions. 
V.C. would like to thank Hideo Kodama, Akihiro Ishibashi and all the
participants of the ExDip2010 workshop in KEK (Tsukuba, Japan) for useful
discussions, and KEK for hospitality while this work was near completion.
This work was supported by the {\it DyBHo--256667} ERC Starting and by FCT -
Portugal through PTDC projects FIS/098025/2008, FIS/098032/2008 and
CTE-AST/098034/2008, CERN projects FP/109306/2009, FP/109290/2009 and by an
allocation through the TeraGrid Advanced Support Program under grant
PHY-090003. E.B.'s and J.R.'s research was supported by NSF grant PHY-0900735.
Computations were performed on the TeraGrid clusters TACC Ranger and NICS
Kraken, the Milipeia cluster in Coimbra, Magerit in Madrid and LRZ in Munich.
%%%%%%%%%%%%%%%%%%%%%%%%%%%%%%%%%%%%%%%%%%%%%%%%%%%%%%%%%%%%%%%%%%%%%%%%%%%%%%%%%%%%%%%%%%%%%%%%%%%%%%%%%

\end{document}